\documentclass[prb,twocolumn,showpacs,amsmath,amssymb]{revtex4}
\usepackage{graphicx}
\usepackage{dcolumn}
\usepackage{bm}
\usepackage{epsfig}
\begin{document}

%\preprint{APS/123-QED}

\title{Magnetic excitations in multiferroic LuMnO$_{3}$ studied by inelastic neutron scattering}

\author{H. J. Lewtas}
\affiliation{
Oxford University Department of Physics, Clarendon Laboratory,  Parks
Road, Oxford, OX1 3PU, UK}

\author{A. T. Boothroyd}
\email{a.boothroyd@physics.ox.ac.uk} \affiliation{ Oxford University
Department of Physics, Clarendon Laboratory,  Parks Road, Oxford,
OX1 3PU, UK}

\author{M. Rotter}
\affiliation{
Oxford University Department of Physics, Clarendon Laboratory,  Parks
Road, Oxford, OX1 3PU, UK}

\author{D. Prabhakaran}
\affiliation{
Oxford University Department of Physics, Clarendon Laboratory,  Parks
Road, Oxford, OX1 3PU, UK}

\author{H. M\"uller}
\affiliation{Institute for Solid State Physics, University of
Technology Vienna, Wiedner Hauptstr 8--10, A-1040 Wien, Austria}

\author{M. D. Le}
\affiliation{Helmholtz-Zentrum Berlin f\"ur Materialien und Energie, Hahn-Meitner-Platz 1, D-14109, Berlin, Germany}

\author{B. Roessli}
\affiliation{
Laboratory for Neutron Scattering, Paul Scherrer Insitute, CH-5232 Villigen PSI, Switzerland
}
\author{J. Gavilano}
\affiliation{
Laboratory for Neutron Scattering, Paul Scherrer Insitute, CH-5232 Villigen PSI, Switzerland
}

\author{P. Bourges.}
\affiliation{
Laboratoire L\'eon Brillouin, CEA-CNRS, CE Saclay, 91191 Gif sur Yvette, France
}

\date{\today}

\begin{abstract}
We present data on the magnetic and magneto-elastic coupling in the
hexagonal multiferroic manganite LuMnO$_3$ from inelastic neutron
scattering, magnetization and thermal expansion measurements. We
measured the magnon dispersion along the main symmetry directions
and used this data to determine the principal exchange parameters
from a spin-wave model. An analysis of the magnetic anisotropy in
terms of the crystal field acting on the Mn is presented. We compare
the results for LuMnO$_3$ with data on other hexagonal RMnO$_3$
compounds.
\end{abstract}

\maketitle

\section{Introduction}

Multiferroic materials have been intensively studied in recent years
following the discovery of compounds that display giant
cross-coupling effects between magnetic and ferroelectric order
parameters\cite{kimura03,hur04,lottermoser04}.  Particular interest
has been aroused by the possibility of new magnetoelectric coupling
mechanisms\cite{mostovoy06,betouras07,cheong07} and the potential
for exploitation in technological
applications\cite{fiebig05,tokura06,eerenstein06}. One of the most
investigated families of multiferroics is the hexagonal manganites
$R$MnO$_3$, which form with $R=$ Sc, Y, Ho, Er, Tm, Yb and Lu. The
magnetoelectric behavior found in this family is associated with
frustrated antiferromagnetic interactions of Mn spins on a
triangular lattice. The compounds formed with $R=$ Sc, Y and Lu are
attractive for fundamental studies because they are not complicated
by additional magnetic contributions from the $R$ ions and because
they form a family in which changes in structure and magnetoelectric
behavior can be correlated in a systematic way. Here we focus on
LuMnO$_3$, and present neutron scattering measurements of the
cooperative magnetic dynamics and measurements of the thermal
expansion by dilatometry. The data provide quantitative information
on the exchange interactions, magnetic anisotropy, and
magnetostriction, all of which play a part in the magnetoelectric
coupling. The results are compared with similar measurements on
YMnO$_3$.

\begin{figure}
\begin{center}
\includegraphics[width=6cm,bbllx=0,bblly=0,bburx=307,bbury=324,angle=0,clip=]{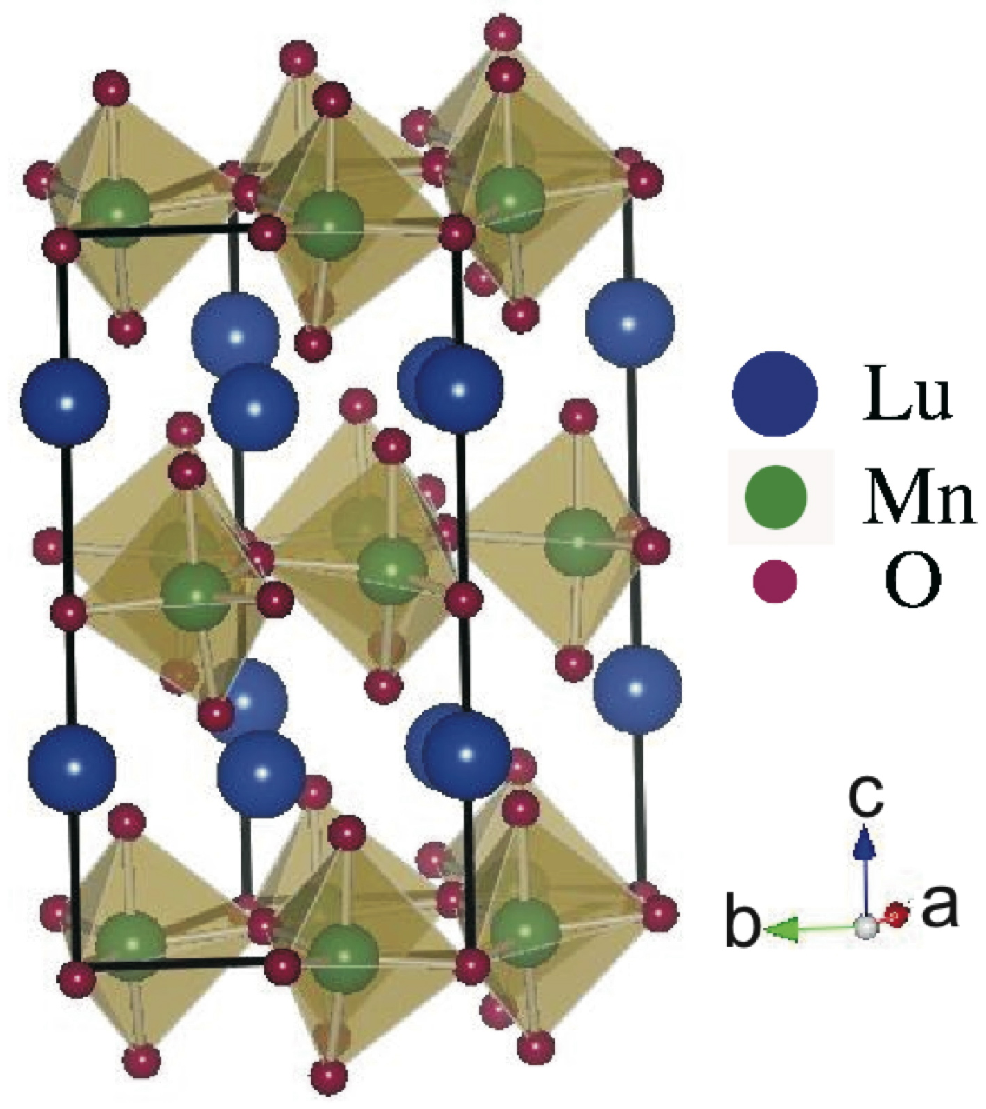}
\includegraphics[width=7cm,bbllx=130,bblly=300,bburx=490,bbury=580,angle=0,clip=]{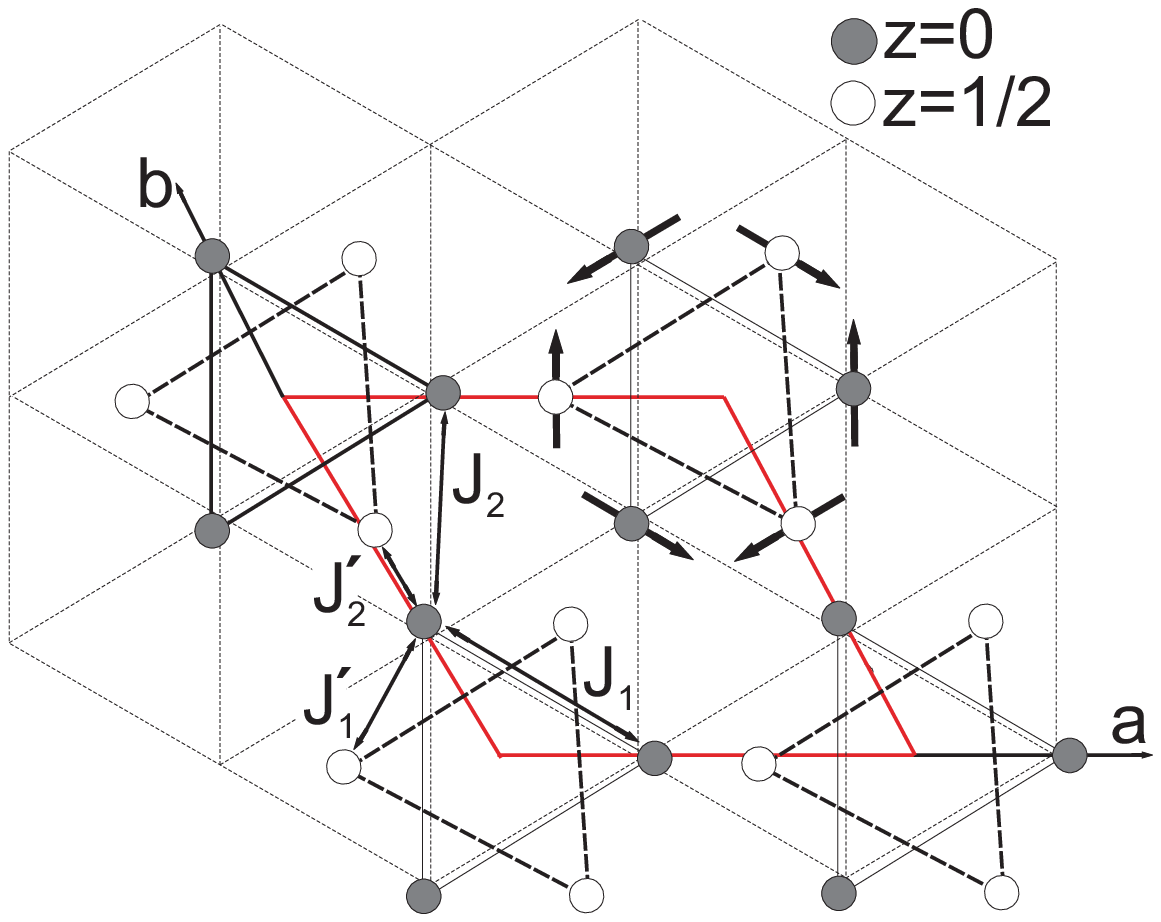}
\caption{(Color online) (a) The hexagonal layered structure of
LuMnO$_3$. (b) Projection of the structure down the $c$ axis showing
the Mn sites. The large (red) diamond is the projection of
chemical/magnetic unit cell, and the filled and empty circles
represent Mn at fractional heights $z=0$ and $z=1/2$, respectively.
The Mn trimerization has been exaggerated for emphasis. The in-plane
($J_1$, $J_2$) and out-of-plane ($J_1'$, $J_2'$) Mn--Mn exchange
couplings used in the spin-wave model are shown. Arrows on the Mn
atoms show the probable magnetic structure of LuMnO$_3$ based on
neutron diffraction\cite{katsufuji02} and optical second harmonic
generation\cite{fiebig00}.} \label{structure}
\end{center}
\end{figure}

The crystal structure of the hexagonal manganites, which is
described by the space group $P6_3cm$, is built from corner-sharing
MnO$_{5}$ bipyramids which form layers parallel to the $ab$ plane
separated by rare earth ions, as shown in Fig.~\ref{structure}. The
Mn ions form a near-ideal triangular lattice. The ferroelectric
distortion, which occurs at a high temperature ($T_{\rm c} >
1000$\,K for LuMnO$_3$, Ref.~\onlinecite{katsufuji02}), is caused by
a tilting of the MnO$_{5}$ bipyramids and a buckling of the $R$
plane, which together create a $\sqrt{3}\times \sqrt{3}$
superlattice distortion (trimerization) of the Mn ions and a
ferroelectric moment along the $c$ axis\cite{katsufuji02,akenNM04}.
The distortion shifts the Mn ions along the $a$ axis away from the
ideal $x=1/3$ position\cite{footnote} --- see Fig.~\ref{structure}.

The magnetic properties of LuMnO$_3$ arise from the (almost)
triangular layers of Mn$^{3+}$ (3$d^4$) ions with $S=2$. Neighboring
spins are coupled by antiferromangetic exchange interactions which
are frustrated by the triangular geometry, as evidenced by the large
ratio of the Weiss to N\'{e}el temperatures $|\Theta/T_{\rm N}| \sim
10$ (Ref.\onlinecite{katsufuji01}), the anomalous magnetic entropy
below $T_{\rm N}$ \cite{katsufuji01}, and the reduction in the value
of the ordered magnetic moment to about 75\% of the full spin-only
moment for $S=2$ (Ref.~\onlinecite{katsufuji02}). The Mn spins in
all the hexagonal $R$MnO$_3$ compounds form a classical
120$^{\circ}$ structure within the triangular layers
(Fig.~\ref{structure})\cite{bertaut63,koehler64,munoz00,fiebig00,brown06,katsufuji02}.
The spins are confined by anisotropy to lie in the $ab$ plane, and
the large inter-layer separation decouples the layers electronically
and makes the magnetism quasi-two-dimensional. In the case of
LuMnO$_3$, magnetic ordering occurs below $T_{\rm N} = 88$\,K.

Evidence for magnetoelectric coupling in $R$MnO$_3$ compounds is
provided by anomalies at $T_{\rm N}$ in the dielectric constant
\cite{huang97,katsufuji01}, lattice
dynamics\cite{souchkov03,jang10,poirier07,petit07}, thermal
conductivity\cite{sharma04} and structural
parameters\cite{akenPRB04,lee05,lee08}. There are also interesting
observations by optical second-harmonic generation which show a
cross-correlation between ferroelectric and magnetic domains due to
the formation of magnetic domain walls below $T_{\rm N}$ which
coincide with ferroelectric domain
walls\cite{fiebig02,goltsev03,hanamura03}.

The precise microscopic mechanism of the magnetoelectric coupling in
$R$MnO$_3$ has not been described yet. Careful structural
measurements on Y$_{1-x}$Lu$_x$MnO$_3$ have shown that an
isostructural transition takes place at $T_{\rm N}$
(Ref.~\onlinecite{lee08}), which causes further displacements of the
ions resulting in a small increase in the ferroelectric
polarization. It was therefore proposed that the magnetoelectric
coupling is driven by a primary magnetoelastic coupling. The origin
of the magnetoelastic coupling, however, remains unclear. One
possibility is that the isostructural distortion may occur in order
to relieve some energy associated with magnetic
frustration\cite{fabreges09}. Another proposal is that the system
might benefit energetically from the Dzyaloshinskii--Moryia
interaction below $T_{\rm N}$ via a small $c$-axis displacement in
the oxygen atoms that bond adjacent Mn atoms\cite{pailhes09}. This
displacement would produce a small additional electric polarization
along the $c$ axis.

\section{Experimental}

Single crystals of LuMnO$_3$ were prepared by the optical
floating-zone technique as follows. Polycrystalline LuMnO$_3$ was
prepared by standard solid-state reaction from high purity
($>$99.999\%) Lu$_2$O$_3$ and MnO$_2$. The polycrystalline powder
was pressed into rods of diameter 8\,mm and length 80\,mm, and
sintered at 1300$^{\circ}$C for 24 hours.  Single crystals were
grown in a four-mirror optical floating-zone furnace (Crystal
Systems Inc.) at a scanning rate of 3--4\,mm\,hr$^{-1}$ with the
feed and seed rods counter-rotating at 30\,rpm.  The growth was
performed in a flowing atmosphere of argon and oxygen in the ratio
12:1. At each stage in the preparation the phase purity of the
product was checked by powder X-ray diffraction.

Unpolarized neutron scattering measurements were performed on a
crystal of mass 1.9\,g on the cold-neutron triple-axis spectrometer
TASP at the SINQ facility (PSI, Switzerland) and on the thermal
triple-axis 2T1 at LLB-Orph\'{e}e (Saclay, France). At TASP, the
crystal was mounted in an `orange' helium cryostat and neutron
spectra were recorded with a fixed final energy of 4.5\,meV. The
corresponding setup at 2T1 was with a closed-cycle refrigerator and
a fixed final energy of 14.7\,meV. Measurements were made with
either $a^{\ast}$ and $c^{\ast}$ or $a^{\ast}$ and $b^{\ast}$ in the
horizontal scattering plane, where $a^{\ast}$, $b^{\ast}$ and
$c^{\ast}$ are the axes of the hexagonal reciprocal lattice. On 2T1
some measurements were also made in the plane parallel to the
$a^{\ast}b^{\ast}$ plane but displaced by 0.5 reciprocal lattice
units along the $c^{\ast}$ axis. The lattice parameters of LuMnO$_3$
referred to the space group $P6_3cm$ are $a = b = 6.05$\,${\rm
\AA}$, $c = 11.4$\,${\rm \AA}$, and the inter-axis angles are
$\alpha = \beta = 90$\,$^{\circ}$, $\gamma = 120$\,$^{\circ}$.
Hence, $a^{\ast} = b^{\ast} = 4\pi/(a\surd 3)$, and $c^{\ast} =
2\pi/c$.

Magnetic measurements were performed with a superconducting quantum
interference device (SQUID) magnetometer on a small piece of crystal
cut from the same rod as the neutron crystal. The thermal expansion
was measured on the same piece of crystal with a miniature
capacitance dilatometer\cite{rotter98-2742,rotter-patent}.

\begin{figure}[ht]
\begin{center}
\epsfig{file=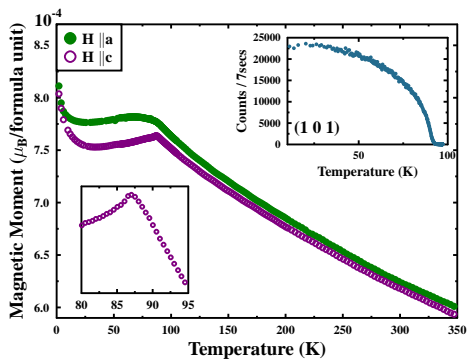,width=\columnwidth}
\caption{(Color online) Zero-field-cooled magnetization measurements
performed in a magnetic field of strength $H = 1000$\,Oe applied
along the $\it{c}$-axis and $\it{a}$-axis. The lower inset shows an
expanded temperature range about the ordering temperature $T_{\rm N}
= 87.5 \pm 0.5$\,K. The upper inset displays the temperature
dependence of the $(101)$ magnetic Bragg peak amplitude.
\label{magnetization}}
\end{center}
\end{figure}

\section{Results}

Figure \ref{magnetization} shows the magnetization of LuMnO$_3$ for
a magnetic field of strength 1000\,Oe applied parallel to the $ab$
plane and along the $c$ axis. Magnetic ordering is signalled by a
sharp peak in the magnetization at $T_{\rm N} = 87.5 \pm 0.5$\,K
(Fig.~\ref{magnetization} lower inset). This is confirmed by the
appearance of magnetic Bragg peaks below $T_{\rm N}$ in neutron
diffraction data (Fig.~\ref{magnetization} upper inset). The
magnetization exhibits a small anisotropy, being slightly larger
when the field is applied parallel to the $ab$ plane ($\chi_{ab}$)
than along the c-axis. This easy-plane anisotropy is consistent with
the observation that the moments lie in the plane in the ordered
phase. The data follows a Curie--Weiss law at high temperatures (not
shown) with a negative Weiss temperature, $\Theta$. From fits of
$1/\chi$ vs $T$ we obtain $\Theta = -819 \pm 2$\,K from $\chi_{ab}$,
and $-837 \pm 1$\,K from $\chi_{c}$. These values are close to those
reported previously from single crystals\cite{katsufuji01}, but
somewhat larger in magnitude than obtained from powder
samples\cite{tomuta01}.  A clear cusp is seen at $T_{\rm N}$ in
$\chi_c$ whereas a broader peak is seen in $\chi_{ab}$. It has been
suggested that the cusp in $\chi_c$ is caused by coupling between
adjacent Mn layers, and the more rounded peak in $\chi_{ab}$ is due
to frustration\cite{katsufuji01}.

\begin{figure}[ht]
\begin{center}
\epsfig{file=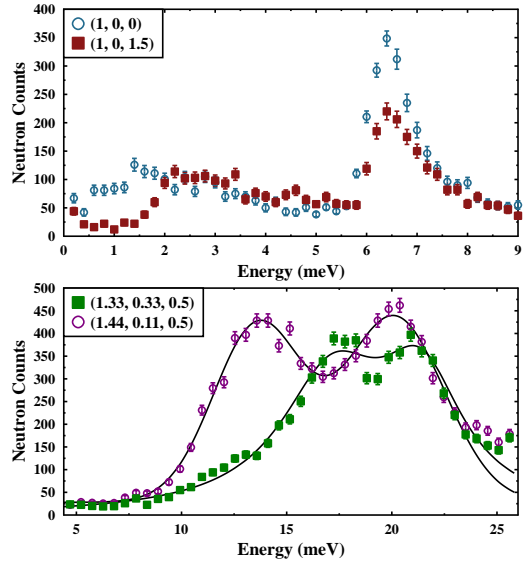,width=\columnwidth} \caption{(Color
online) Neutron inelastic scattering from LuMnO$_3$ measured on (a)
TASP at $T=5$\,K and (b) 2T1 at $T=13.5$\,K. The data are from
constant-wavevector scans at the indicated positions in reciprocal
space. The lines in (b) are fits to a lineshape comprising two
gaussian functions. \label{spectrum}}
\end{center}
\end{figure}

Examples of neutron scattering spectra from both instruments are
presented in Fig.~\ref{spectrum}. Figure~\ref{spectrum}(a) shows
energy scans recorded on TASP at the scattering vectors ${\bf Q} =
(1,0,0)$ and $(1,0,1.5)$, both of which contain two asymmetric
peaks.  Figure~\ref{spectrum}(b) shows data at ${\bf Q} =
(1.33,0.33,0.5)$ and $(1.44,0.11,0.5)$ measured on 2T1. Since the ordered moment on the Mn sites is
relatively large ($\sim 3$\,$\mu_{\rm B}$) the scattering from
magnons is expected to be much stronger than phonon scattering at
these relative small scattering vectors. This, together with the
resemblance of the spectra to previous measurements on YMnO$_3$,
gives us confidence that the main features in the spectra correspond
to magnon excitations.

\begin{figure}[ht]
\begin{center}
\epsfig{file=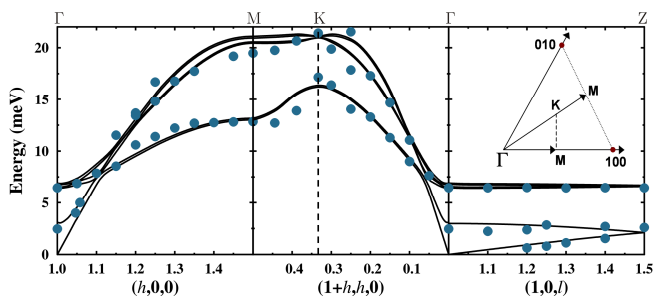,width=\columnwidth} \caption{(Color
online) Spin-wave dispersion of LuMnO$_3$. Solid circles are the
peak centers obtained from fits to scans such as those shown in
Fig.~\ref{spectrum}. Solid lines are calculated from the model
described in the text with parameters $J_{1} = -4.09(2)$\,meV,
$J_{2} = -1.54(5)$\,meV, $J_{2}' = +0.019(2)$\,meV, $J_{1}' = 0$,
and $D = -0.48$\,meV. The inset is a sketch of the $(h,k,0)$ plane
in reciprocal space showing the path $\Gamma$MK$\Gamma$.
\label{dispersion}}
\end{center}
\end{figure}

To determine the magnon dispersion we fitted the peaks with Gaussian
or Lorentzian functions (depending on the peak shape) on a linear
background. The fitted peak positions have been collected together
into a dispersion curve which is plotted in Fig.~\ref{dispersion},
including data from both TASP and 2T1. Measurements on TASP extended
from below 1\,meV up to about 13meV, while measurements at 2T1
covered the range from about 5\,meV up to the energy of the highest
modes. Measurements were performed in several different zones to
find the maximum intensity and to check that the peaks were periodic
in reciprocal space.

\begin{figure}[ht]
\begin{center}
\epsfig{file=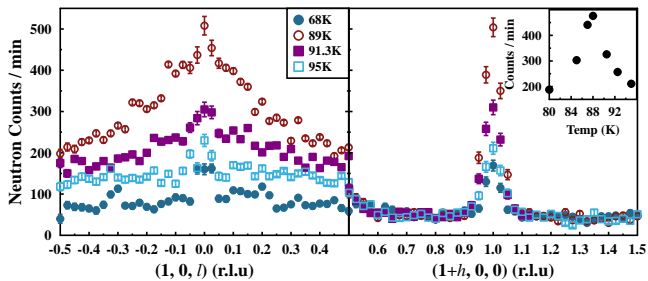,width=\columnwidth}
\caption{(Color online) Diffuse neutron scattering around the
forbidden nuclear reflection $(100)$. The inset shows the
temperature variation in the diffuse scattering intensity at
$(1,0,0)$. \label{diffuse}}
\end{center}
\end{figure}

Figure~\ref{diffuse} shows diffuse scattering measurements in the
vicinity of the point $(1,0,0)$ in reciprocal space at temperatures
close to $T_{\rm N}$. The $(100)$ reflection has zero nuclear
structure factor, and we observed zero magnetic intensity here at
low temperatures. The scans in Fig.~\ref{diffuse} reveal strong
diffuse scattering at temperatures close to $T_{\rm N}=87.5 \pm 0.5$
with maximum intensity at $T_{\rm N}$ itself, as shown in the inset.
The diffuse scattering is highly anisotropic, being very broad in
the $(0,0,l)$ direction but much sharper in the $(h,0,0)$ direction.

\begin{figure}[ht]
\begin{center}
\epsfig{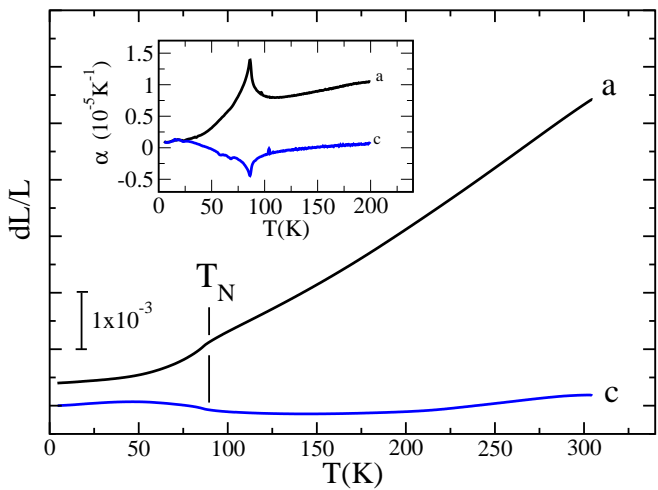} \caption{(Color online)
Thermal Expansion of LuMnO$_3$ measured along the hexagonal $a$- and
$c$ directions. The inset shows the thermal expansion coefficient
$\alpha$ as determined from the temperature derivative of the strain
${\rm d}L/L$.\label{the}} \end{center}
\end{figure}

Figure~\ref{the} shows measurements of the thermal expansion of
LuMnO$_3$ parallel to the $a$ and $c$ axes.  A magnetoelastic
anomaly is clearly visible at the N\'eel temperature, both in the
strain ${\rm d}L/L$ and in its temperature derivative $\alpha =
L^{-1}{\rm d}L/{\rm d}T$. On cooling through $T_{\rm N}$ the
magnetoelastic strain expands the $c$ axis and shrinks the hexagonal
plane. As a check, we also measured the thermal expansion in the
hexagonal plane in the direction normal to $a$. The data resemble
the behaviour of $da/a$ to within the experimental error. Above
$T_{\rm N}$ the thermal expansion is highly anisotropic. The
$c$-axis strain is almost temperature independent.

\section{Analysis and discussion}

We first review the magnetic structure of the $R$MnO$_3$ compounds \cite{bertaut63,koehler64,munoz00,fiebig00,brown06},
with particular reference to LuMnO$_3$. The six Mn sites in the unit
cell (Fig.~\ref{structure}) form two near-equilateral triangles, one
in the $z=0$ layer and the other in the $z=1/2$ layer. The spins on
these triangles lie in the basal plane and order in 120$^{\circ}$
structures. Symmetry constrains the relation between the $z=0$ and
$z=1/2$ layers to two possibilities, conventionally labeled $\alpha$
and $\beta$. In the $\alpha$ structure the spin on the Mn at
$(x,0,0)$ ($x \approx 1/3$) is parallel to that at ($1-x,0,1/2)$,
whereas in the $\beta$ structure the spins on these two sites point
in opposite directions. Each spin makes an angle $\psi$ to the unit
cell axis on which it lies, and the magnetic structure factors
depend on $\psi$ and on the stacking relation ($\alpha$ or $\beta$).
For $x=1/3$, the magnetic structures occur in homometric pairs, such
that the magnetic diffraction intensities for the configuration
($\alpha$, $\psi$) are identical with those from ($\beta$, $\psi \pm
90^{\circ}$).

In the case of LuMnO$_3$, Katsufuji {\it et al.} concluded from
neutron powder diffraction measurements\cite{katsufuji02} that the
low temperature structure is one of two possibilities, either
($\alpha$, $\psi = \pm 90^{\circ}$) or ($\beta$, $\psi =0^{\circ}$
or $180^{\circ}$). These two structures transform respectively like
the $\Gamma_4$ and $\Gamma_2$ irreducible representations of the
space group $P6_3cm$ (Ref.~\onlinecite{munoz00}). Only this
particular homometric pair have a completely absent $(100)$ magnetic
reflection. In our single crystal measurements we also found very little intensity at the $(100)$ reflection (but relatively strong
intensity for the $(101)$ reflection
--- see the inset in Fig.~\ref{magnetization}) at the lowest temperature
($T=2$\,K), in agreement with Katsufuji {\it et al.} The homometric
pairs can in principle be distinguished by optical second harmonic
generation (SHG). Using this method, Fiebig {\it et al.} found that
their sample of LuMnO$_3$ was a two-phase mixture of $\alpha$
structures with $\psi = 0^{\circ}$ and an unspecified other $\psi$
value\cite{fiebig00}. Since our low-temperature data, as well as
that of Ref.~\onlinecite{munoz00}, conclusively rule out any
$\alpha$ structure which does not have $\psi \simeq \pm 90^{\circ}$
it is difficult to see how to reconcile the diffraction and SHG
results. For the purpose of modeling the spin wave spectrum we will
assume the ($\alpha$, $\psi = \pm 90^{\circ}$) structure, as shown
in Fig.~\ref{structure}. The magnetic spectrum of this and its
homometric partner are not distinguishable at the level of precision
of our data.

We calculated the spin wave spectrum from the spin Hamiltonian
\begin{equation}{\label{eq:spinwave}}
{\mathcal H} = -\sum_{\left \langle ij \right \rangle} J_{ij} {\bf
S}_{i}\cdot {\bf S}_{j}-D\sum_{i} \left ( S^z_i \right )^2,
\end{equation}
with two in-plane near-neighbor interactions ($J_1$ and $J_2$) and
two inter-plane interactions ($J_{1}'$ and $J_{2}'$) defined as
shown in Fig.~\ref{structure}. The first summation in
(\ref{eq:spinwave}) is over pairs of spins with each pair counted
once so that the $J$ constants are per spin pair. The second term
models the out-of-plane anisotropy with a single-ion anisotropy
parameter $D$. We neglect the small in-plane anisotropy since the
in-plane magnon gap was too small to measure in our experiment.

Analytic expressions have been given previously for the spin-wave
energies derived from spin Hamiltonians similar to
(\ref{eq:spinwave}),
Refs.~\onlinecite{sikora88,sato03,vajk05,chatterji07}. These
expressions have been obtained via the usual transformation of the
Hamiltonian into a quadratic form of boson normal-mode operators in
the linear approximation. Here we use an alternative method based on
dynamical matrix diagonalisation (DMD) as outlined in previous work
\cite{rotter06-400}, which is implemented in the McPhase software
package\cite{rotter-mcphase}. This formulation employs the random
phase approximation to calculate the magnon cross sections in
addition to the dispersion relations.

There are six Mn spins per unit cell, which gives rise to a total of
six spin-wave modes for each wavevector. As the interlayer coupling
is small, the in-plane dispersion relations appear as three branches
each containing two nearly-degenerate modes. The degeneracy of the lowest two
modes is lifted close to the $\Gamma$ point and along $\Gamma$Z by
the effect of the $J_1'$ and $J_2'$ interactions, while the upper
four modes are almost degenerate along $\Gamma$Z. This degeneracy
precludes the possibility to fit accurate values for $J_1'$ and
$J_2'$ independently, and so we chose to fix $J_1' = 0$ and to vary
$J_2'$ under the constraint that $J_2'
> 0$ to maintain the stability of the $\alpha$ structure.

A least-squares fitting procedure returned the following values for
the model parameters: $J_{1} = -4.09(2)$\,meV, $J_{2} =
-1.54(5)$\,meV, $J_{2}' = +0.019(2)$\,meV ($J_{1}' = 0$), and $D =
-0.48$\,meV.  The calculated dispersion relations from the model
with these parameters are shown in Fig.~\ref{dispersion}. The
agreement is seen to be very good, and the parameters are well
constrained by the data. For example, the 6.5\,meV gap to the upper
mode at $\Gamma$ is sensitive to the single-ion anisotropy, and
$J_{2}'$ controls the dispersion in the out-of-plane direction. The
splitting of the magnon peaks in the vicinity of the $K$ point, seen
in Fig.~\ref{spectrum}, is sensitive to the difference between $J_{1}$
and $J_{2}$. Qualitative agreement between the measured and
calculated magnon cross sections gave us further support for the
obtained parameters.

As a check, we calculated the bulk magnetization using the best-fit
exchange and anisotropy parameters. The $T_N$ predicted by the
mean-field model is about 2.5 times larger than the observed $T_N$,
presumably as a consequence of frustration. The calculated
susceptibility has a small easy-plane anisotropy consistent with the
measured susceptibility, Fig.~\ref{magnetization}. The magnetization
of the sample as a function of applied field (not shown) is linear
and almost identical in the $\it{a}$ and $\it{c}$ directions.
Increasing the single-ion anisotropy in the model creates a step in
the $a$-axis magnetization which is not observed. This adds to the
evidence that the single-ion anisotropy is very small compared to
the exchange interactions.

The exchange parameters obtained here show that the dominant
magnetic interaction is the in-plane antiferromagnetic superexchange
via the $\sim$$120^{\circ}$ Mn-O-Mn path. The inter-layer
superexchange is two orders of magnitude weaker, confirming that the
magnetism in LuMnO$_3$ is highly two-dimensional. It is interesting
to compare the magnetic spectrum of LuMnO$_3$ investigated here with
those obtained from similar measurements on YMnO$_3$
(Refs.~\onlinecite{sato03,chatterji07,petit07,fabreges09}) and
HoMnO$_3$ (Ref.~\onlinecite{vajk05}). Qualitatively, the spectra of
the three compounds look very similar, but the overall band width of
the LuMnO$_3$ spectrum is about 30\% larger than that of YMnO$_3$
and HoMnO$_3$ (21\,meV compared with 16\,meV). Consistent with this,
the fitted exchange parameters for LuMnO$_3$ are found to be
systematically larger than those of YMnO$_3$ and HoMnO$_3$. This
accounts for the difference in the antiferromagnetic ordering
temperatures of these compounds: $T_{\rm N} \approx 88$\,K
(LuMnO$_3$) compared with $T_{\rm N} \approx 72$\,K (YMnO$_3$) and
$T_{\rm N} \approx 75$\,K (HoMnO$_3$) and the larger Weiss
temperature of LuMnO$_3$ ($\Theta \approx -850$\,K) compared with
YMnO$_3$ ($\Theta \approx -700$\,K)\cite{katsufuji01}.

The stronger magnetic interactions in LuMnO$_3$ fits with the
systematic trend in the ionic radii and the lattice
parameters\cite{katsufuji02}, i.e. the smaller the ionic radius the
smaller the unit cell and the stronger the exchange interactions.
The single-ion anisotropy parameter $D$ is also found to be larger
for LuMnO$_3$ ($D=-0.48$\,meV) than for YMnO$_3$ ($D = -0.28$ to
$-0.33$\,meV) and HoMnO$_3$ ($D = -0.38$\,meV). This could be
another consequence of the small differences in the structural
parameters of these compounds.

The character of the diffuse scattering from LuMnO$_3$ close to
$T_{\rm N}$ (Fig.~\ref{diffuse}) strongly resembles that observed
from YMnO$_3$.\cite{roessli05} The appearance of scattering which is
very broad along $c$ but relatively sharp in the plane indicates
that the inter-layer correlations are very weak, consistent with the
small $J_1'$ and $J_2'$ and two-dimensional nature of the magnetic
system. The strong enhancement in the diffuse scattering intensity
around $(1,0,0)$ was also observed in powder diffaction measurements
on LuMnO$_3$.\cite{katsufuji02} These showed that the diffuse peak
persists up to at least $\sim 3T_N$,\cite{katsufuji02} which was
interpreted as evidence for strong geometric frustration.

Careful powder diffraction measurements\cite{lee08} have shown that
the magnetically-induced ferroelectricity in $R$MnO$_3$ is
associated with an isostructural transition involving an additional
rotation of the MnO$_5$ bipyramids, and that the increase in
ferroelectric polarization below $T_{\rm N}$ scales with the square
of the ordered moment. The thermal expansion of LuMnO$_3$ reported
here (Fig.~\ref{the}) reveals a striking magnetoelastic anomaly at
$T_{\rm N}$, consistent with the diffraction data of Lee {\it et
al.} (Ref.~\onlinecite{lee08}) who argued that magnetoelastic
coupling (exchange striction) is the primary source of the
magnetoelectric coupling\cite{lee08}.  One might expect, therefore,
that the magnetoelectric effect would scale with the strength of the
exchange interactions and hence be greater in LuMnO$_3$ than in
YMnO$_3$. Support for this idea is provided by the
magnetically-induced polarization calculated from the measured
atomic displacements below $T_{\rm N}$, which indeed appears to be
systematically larger for LuMnO$_3$ than for YMnO$_3$
(Ref.~\onlinecite{lee08}). However, given the large experimental
uncertainties in the values of the small atomic displacements this
evidence should be considered tentative.

Another notable feature of the thermal expansion is how small the
$c$-axis thermal expansion is relative to the $ab$ plane expansion
above $T_{\rm N}$ (see Fig.~\ref{the}). This effect, which is
observed both for LuMnO$_3$ and YMnO$_3$, does not correlate with
the compressibility of these materials, which is similar along the
$c$ direction and in the $ab$ plane\cite{kozlenko_JEPT}. The
relatively isotropic compressibility suggests that the anomalous
$c$-axis thermal expansion is not due to a straightforward
anharmonicity in the interatomic potentials along the $c$ axis, and
it would be interesting to find out what is responsible for it.

\begin{figure}[ht]
\begin{center}
\epsfig{file=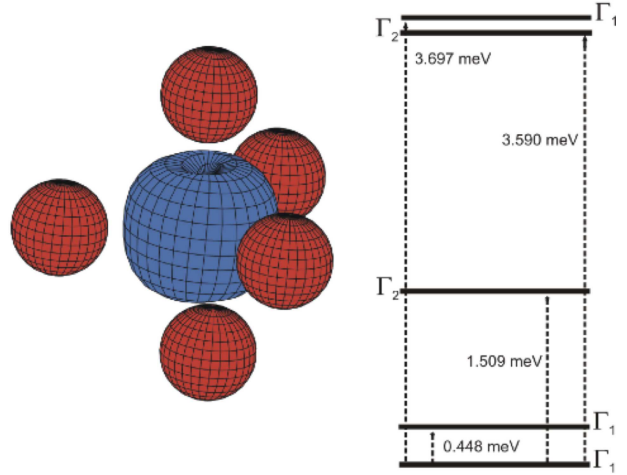,width=\columnwidth} \caption{Left:
Calculation of the thermally-averaged charge density in the MnO$_5$
bipyramid of LuMnO$_3$ at $T=4$\,K. The 3$d$ charge is represented
(in blue) by a surface of constant charge density obtained from a
calculation of the crystal field acting on Mn$^{3+}$ assuming point
charges of $-2|e|$ on each of the five nearest oxygen neighbors
(shown in red). Right: Low-lying energy levels of Mn$^{3+}$ split by
the point-charge crystal field via spin--orbit coupling. The
symmetry of each level is labeled according to the irreducible
representation of the point group $C_s$, which describes the local
symmetry around the Mn site. \label{point charge}}
\end{center}
\end{figure}

Before concluding, we investigate the origin of the small magnetic
anisotropy of LuMnO$_3$, which is represented in the Hamiltonian
(\ref{eq:spinwave}) by the phenomenological $D(S^z)^2$ term. For
reference, we performed a point-charge calculation of the the
crystal field at the Mn sites assuming Mn to be in the Mn$^{3+}$
state with $d^4$ configuration and using the structural parameters
reported by Katsufuji {\it et al.} in Ref.~\onlinecite{katsufuji02}.
We included only the five nearest oxygen neighbors of Mn in the
MnO$_5$ bipyramid, as shown in Fig.~\ref{point charge}. The
ground-state $S=2$ manifold is split by the crystal field via the
spin--orbit interaction. The calculated splitting, due to the crystal field alone, is also shown in
Fig.~\ref{point charge}. This model of the crystal field predicts
that the ordered magnetic moment points along the normal to the
local mirror plane, as shown in Fig.~\ref{structure}, and with the inclusion of the exchange field from the neighboring Mn ions predicts
an anisotropy gap in the magnon spectrum of $\sim$10\,meV. This size
of gap is in clear disagreement with the observed spin-wave modes,
which have a gap of less than 1\,meV at the zone centre,
Fig.~\ref{dispersion}. The anisotropy gap may be reduced in the
model if the local symmetry is increased to $C_{3v}$, i.e. by
lessening the degree of Mn trimerization and tilting of the MnO$_5$
bipyramid. We conclude, therefore, that the single-ion anisotropy is
controlled by a tiny distortion of the ideal MnO$_5$ bipyramid and
that the anisotropy is much smaller than that predicted by a simple
point-charge model.

\section{Conclusions}

We have measured the magnon dispersion in LuMnO$_3$ and achieved a
very good description of the spectrum using the Heisenberg
Hamiltonian, Eq.~(\ref{eq:spinwave}). We also observed a striking
magnetoelastic coupling at $T_{\rm N}$ in the thermal expansion. The
results are qualitatively similar to those previously obtained on
the sister compound YMnO$_3$. The bandwidth of the one-magnon
spectrum of LuMnO$_3$ is about 30\% larger than that of YMnO$_3$,
and the difference between the two nearest-neighbour in-plane
exchange constants $J_1$ and $J_2$ is greater for LuMnO$_3$ than for
YMnO$_3$. As the magnetic interactions are stronger in LuMnO$_3$
than in YMnO$_3$ we expect the magnetically-induced ferroelectric
polarization to be greater in LuMnO$_3$. The available diffraction
data provides tentative support for this.

\section{Acknowledgments}
This work was performed partly at the Swiss spallation neutron
source SINQ, Paul Scherrer Institut, Switzerland, and partly at the
Laboratoire L\'{e}on Brillouin, Saclay, France. We are grateful for
support from by the Engineering and Physical Sciences Research
Council of Great Britain and from the European Commission under the
7th Framework Programme through the `Research Infrastructures'
action of the `Capacities' Programme, Contract No: CP-CSA
INFRA-2008-1.1.1 Number 226507-NMI3.

\end{document}